# Reconciling interoperability with efficient Verification and Validation within open source simulation environments[⋆]

Stefano Sinisi[∗], Vadim Alimguzhin, Toni Mancini, Enrico Tronci

*Computer Science Department, Sapienza University of Rome, Italy*

**Abstract**

A Cyber-Physical System (CPS) comprises physical as well as software subsystems. Simulation-based approaches are typically used to support design and Verification and Validation (V&V) of CPSs in several domains such as: aerospace, defence, automotive, smart grid and healthcare.

Accordingly, many simulation-based tools are available to support CPS design. This, on one side, enables designers to choose the toolchain that best suits their needs, on the other side poses huge interoperability challenges when one needs to simulate CPSs whose subsystems have been designed and modelled using different toolchains. To overcome such an interoperability problem, in 2010 the Functional Mock-up Interface (FMI) has been proposed as an open standard to support both Model Exchange (ME) and Co-Simulation (CS) of simulation models created with different toolchains. FMI has been adopted by several modelling and simulation environments. Models adhering to such a standard are called Functional Mock-up Units (FMUs). Indeed FMUs play an essential role in defining complex CPSs through, *e.g.*, the System Structure and Parametrization (SSP) standard.

Simulation-based V&V of CPSs typically requires exploring different simulation scenarios (*i.e.*, exogenous input sequences to the CPS under design). Many such scenarios have a shared prefix. Accordingly, to avoid simulating many times such shared prefixes, the simulator state at the end of a shared prefix is saved and then restored and used as a start state for the simulation of the next scenario. In this context, an important FMI feature is the capability to save and restore the internal FMU state on demand. This is crucial to increase efficiency of simulation-based V&V. Unfortunately, the implementation of this feature is not mandatory and it is available only within some commercial software. As a result, the interoperability enabled by the FMI standard cannot be fully exploited for V&V when using open-source simulation environments.

---

This motivates developing such a feature for open-source CPS simulation environments.

Accordingly, in this paper, we focus on JModelica, an open-source modelling and simulation environment for CPSs based on an open standard modelling language, namely Modelica. We describe how we have endowed JModelica with our open-source implementation of the FMI 2.0 functions needed to save and restore internal states of FMUs for ME. Furthermore, we present experimental results evaluating, through 934 benchmark models, correctness and efficiency of our extended JModelica. Our experimental results show that simulation-based V&V is, on average, 22 times faster with our get/set functionality than without it.

## 1. Introduction

Cyber-Physical Systems (CPSs) integrate physical (*e.g.*, mechanical, electrical, etc.) and software (*e.g.*, Software Digital Radios, SDRs, control software, etc.) subsystems. CPSs are widely used in several fields like aerospace, smart grid, manufacturing, automotive, robotics, and health-care (see, *e.g.*, [1, 2, 3, 4, 5, 6, 7, 8, 9]). As they couple the discrete and continuous dynamics of software and physical subsystems, respectively, CPSs are typically defined by means of hybrid systems (see, *e.g.*, [10]). Due to such a complex nature of CPS models, simulation-based approaches are *typically* used to support design and Verification and Validation (V&V) activities. V&V aims at checking whether the CPS model behaviour satisfies given specifications, *e.g.*, safety properties (see, *e.g.*, [11, 12]). In the literature, there are *many* examples where V&V activity is performed by means of numerical simulations (see, *e.g.*, [13, 14, 15, 16, 17, 18, 19] and [20] for a survey).

Many simulation-based software tools are available to support CPS design such as, *e.g.*, AUTOSAR, Automation Studio, AVL Cruise, CATIA, ControlBuild, Simulink, dSpace, EnergyPlus, IBM Rational Rhapsody, ICOS, IGNITE, Dymola, JModelica, MapleSim, Ptolemy II and Virtual Engine. On one side, the increasing availability of those software tools enables designers to choose the tool chain that best suits their needs. On the other side, such an availability poses huge interoperability (model exchange) and integration (co-simulation) challenges between CPSs modelled using different languages and/or tools.

Furthermore, since physical and/or software subsystems are *usually* designed by different companies (*e.g.*, Original Equipment Manufacturers, OEMs), it is also crucial to preserve Intellectual Property (IP) (see, *e.g.*, [21, 22, 23]). As a result, even carrying out simulations of those models may pose problems.

To overcome such interoperability problems, a standardised format, namely, Functional Mock-up Interface (FMI), has been proposed in 2010 as an open standard. Models adhering to FMI are called Functional Mock-up Units (FMUs).

The current version, *i.e.*, FMI 2.0, enables both Model Exchange (ME) and Co-Simulation (CS). ME refers to, *e.g.*, the usage of FMUs within different simulation environments, while CS refers to, *e.g.*, the distributed simulation of heterogeneous systems coupling together several FMUs. Recently, another standard has been defined, *i.e.*, System Structure and Parametrization (SSP), to describe relationships among systems of interconnected FMUs and their parametrisation in order to be used in different simulation environments [24].

Typically, simulation-based V&V of CPSs requires exploring different simulation scenarios, *i.e.*, sequences of controllable and/or uncontrollable exogenous inputs. In this setting, to avoid simulating many times a common prefix of different scenarios, the simulator state is saved in order to be restored later as a start state. This improves efficiency of simulation-based V&V approaches by simulating only once the same prefix of different simulation scenarios (see, *e.g.*, [25] and citations thereof).

Not only V&V approaches benefit from such an optimization, but also, *e.g.*, approaches for the automated synthesis of plans/control strategies such as [26, 27, 28, 29] where simulation-based Model Predictive Control (MPC) methods are used and [30, 31] where a pharmacological treatment strategy is automatically designed by means of simulations.

To this end, modern simulators (such as, *e.g.*, Simulink [32]) offer their own Application Programming Interface (API) allowing to save and restore the internal simulator state on demand. As well as the other simulators, the FMI 2.0 API specifies a way of saving and restoring internal FMU states. This feature is crucial to increase efficiency of simulation-based V&V.

Unfortunately, the implementation of such a feature is not mandatory according to FMI 2.0 specifications. Hence, even if the FMI standard is currently adopted by several modelling environments (see [33] for a full list) only a few commercial software implement this feature within their generated FMUs.

Among them we note Dymola, a state-of-the-art modelling and simulation environment [34], that is based on Modelica [35], an open-standard language for modelling dynamical systems.

Modelica is an object-oriented, equation-based language for model-based development. Moreover, it allows the definition of complex dynamical systems as a network of smaller subsystems. Currently, Modelica is widely used by many companies and it is adopted by several simulation environments: commercial (*e.g.*, SimulationX [36] and SystemModeler [37]) as well as open source (*e.g.*, Open Modelica [38] and JModelica [39]). However, none of the currently available open-source Modelica environments implement the FMI 2.0 optional feature for saving and restoring FMU states. As a result, the interoperability enabled by the FMI standard cannot be fully exploited for V&V when using those open-source environments.

This motivates developing such a feature for Modelica-based and open-source CPS simulation environments.

In the literature the save-and-restore feature has been widely exploited for formal verification both for finite state systems (see, *e.g.*, [40, 41, 42, 43]) as well as, in a simulation-based framework, for CPSs (see, *e.g.*, [44, 45, 25]). We refer the readers to those references and citations therein for general considerations and more details about algorithms that exploit such a feature.

In this paper, we provide methods and tools to implement FMI 2.0 functionalities that save and restore the internal FMU state for ME FMUs and we focus on JModelica modelling and simulation environment.

Furthermore, we present experimental results to evaluate the correctness of our proposed implementation. Finally, we also conduct an analysis of performance focusing on V&V approaches that drive given FMUs by means of simulations. To do so, we analyse 934 FMUs generated from benchmarks models taken from widely-used repositories and we show that, using our tool, a V&V activity in the style of [44] is, on average, 22 times faster than without it.

We remark that such a proposed implementation closes an important gap between commercial and open-source Modelica environments. Indeed, it enables the application of the above-mentioned simulation-based V&V approaches to CPSs modelled within open-source Modelica environments. Furthermore, it also fosters the development of new approaches.

Moreover, the aim of this paper is also to encourage developers of open-source simulation environments to implement modern functionalities of current FMI specifications; and to push towards future FMI specifications that include a mandatory implementation of such functionalities. For these reasons, our proposed implementation is free and *publicly available* at the following repository: https://bitbucket.org/mclab/jmodelica.org, as a fork of the JModelica 2.1 open-source distribution, which is developed by Modelon[1] and currently available upon request.

The paper is organised as follows. Section 2 discusses related work. Section 3 describes our methodology to implement get/set FMU state functionality and to evaluate correctness and performance of the proposed implementation. Section 4 presents experimental results. Finally, Section 5 draws conclusions.

## 2. Related Work

The Functional Mock-up Interface (FMI) is a widely used standard for model-based design and analysis of Cyber-Physical Systems (CPSs). In the literature we find many examples in different fields (see, *e.g.*, [46, 47, 48, 49]). Both Co-Simulation (CS) and Model Exchange (ME) paradigms are crucial to handle large-scale systems (see, *e.g.*, [50, 51]). The former is needed to integrate and analyse in a distributed fashion different subsystems having different characteristics as, *e.g.*, the embedded solver needed to simulate the model (see, *e.g.*, [52, 53] for recent surveys). The latter is used to, *e.g.*, import and/or export models as black-box objects among different simulation environments in order to analyse and simulate them with different solvers. As anticipated in Section 1, we focus on Modelica-based open-source platforms compliant with the last version of FMI for ME, namely 2.0.

---

[1] https://www.modelon.com

| Tool | Modelica Support | Get/Set FMU state functionality |
|---|:---:|:---:|
| DACOSSIM | - | - |
| FMI4J | - | - |
| QTronic FMUSDK | - | - |
| JModelica | ● | - |
| OpenModelica | ● | - |
| PythonFMU | - | - |
| Simulix | - | - |
| Reference FMUs | - | ME & CS |
| Our extended JModelica | ● | ME |

Table 1: Open source modelling and simulation environments supporting FMI 2.0.

Several commercial and open-source modelling and simulation environments support FMI 2.0 (more than 100, see [33] for a full list).

Among commercial environments, we note Matlab/Simulink [32]: a widely-used and well-known platform for model-based design. It also offers its own Application Programming Interface (API) for advancing forward and backward the simulation by rolling back simulator states (thanks to a save/restore functionality). However, being a commercial software, Simulink supports only export of Functional Mock-up Units (FMUs) for CS and does not implement get/set functionality of FMI 2.0.

The FMI standard is supported also by some major Modelica-based commercial modelling and simulation environments such as:

1. Dymola 2020 [34], which, to the best of our knowledge, is the only Modelica-based modelling and simulation platform *fully* implementing FMI 2.0 standard. Hence, it generates FMUs for CS and also for ME having get/set state functionality implemented.
2. SystemModeler 12 [37], which allows users to generate their models as FMUs for both CS and ME. However, no FMU get/set state functionality is implemented.
3. SimulationX 4.1 [36], which also provides an implementation to get/set state functionality. However such a functionality is enabled only within FMUs for CS.

As motivated in Section 1, we focus on open-source modelling and simulation environments.

In Table 1 we compared several open-source platforms which support FMI 2.0. We split these platforms in the following groups:

1. The first group consists of all modelling and simulation environments supporting Modelica. Among those, we note OpenModelica [38] developed by the Open Source Modelica Consortium. OpenModelica translates Modelica models into C code, which, in turn, is compiled in a black-box executable format. Only upon user request it is also possible to wrap such a format into an FMU. Conversely, JModelica is FMI-based. Indeed, it directly generates FMUs from the input Modelica code. Hence, FMUs are first-class citizens. This is the main reason why we focus on JModelica. However, we would like to clarify that, with a little effort, our proposed methodology can be quickly adapted to OpenModelica as well.
2. The second group consists of all modelling and simulation environments which support FMI 2.0 but are not based on Modelica. Among those, *Reference FMUs* [54] implements get and set of FMU states. However, besides the fact that it is not Modelica-based, this software is just a small tool used for debugging FMUs.

A work close to ours is the NANDRAND simulation platform [55] which provides and simulates ready-to-use models to measure energy performance of physical buildings. Such models can be exported as FMUs compliant to FMI 2.0 and support get/set state functionality. It is worth noting that NANDRAND imports external FMUs (*e.g.*, generated by a Modelica-based environment) and couples them to NANDRAND models in a CS setting. However, as the authors specify, such a use-case works if and only if the imported FMUs support state saving and restoring.

Another tool related to our work is the FMI++ library [56] which is useful to simulate FMUs for ME. One of the features implemented by this library is a wrapper class, called `RollbackFMU`. Such a class stores, at each integration step, the current state of the input FMU in order to enable a rollback functionality. The functions used to save and restore the state are `fmi2GetContinuousState` and `fmi2SetContinuousState`, respectively. However, as FMI specifications clarify, such functions are not intended to set and get the complete FMU state but just values of continuous variables of the given FMU. As Section 3.2 describes in details, this is not enough. The state of an FMU has a complex structure where values of continuous model variables are just a small fraction of it.

## 3. Methods

In this section, first, we introduce Functional Mock-up Interface (FMI) 2.0 for Model Exchange (ME) with a particular emphasis on JModelica Functional Mock-up Unit (FMU) implementation. Second, we describe the steps needed to enable saving and restoring of internal FMU states. Finally, we present our approach to evaluate the correctness of our implementation and to measure performance when adopting such implementation within simulation-based Verification and Validation (V&V) methods.

*3.1. FMI 2.0 for Model Exchange*

FMI ME 2.0 [57] specifies an interface to the model of a dynamical system defined by differential, algebraic and/or difference equations. An executable im-plementing FMI is called FMU. Any simulation environment supporting FMI can simulate an FMU produced by any modelling environment. In particular, an FMU is a ZIP archive consisting of the following parts. First, C code and/or binaries implementing functions defined by the FMI Application Programming Interface (API). Second, an Extensible Markup Language (XML) document with model metadata including, *e.g.*, model identifier, names of variables and capability flags. The latter allows omitting implementation of non-mandatory functionalities. For example, the `canGetAndSetFMUstate` and `canSerializeFMUstate` capability flags indicate whether FMU supports functionality to save/restore and serialize/deserialize its internal state or not.

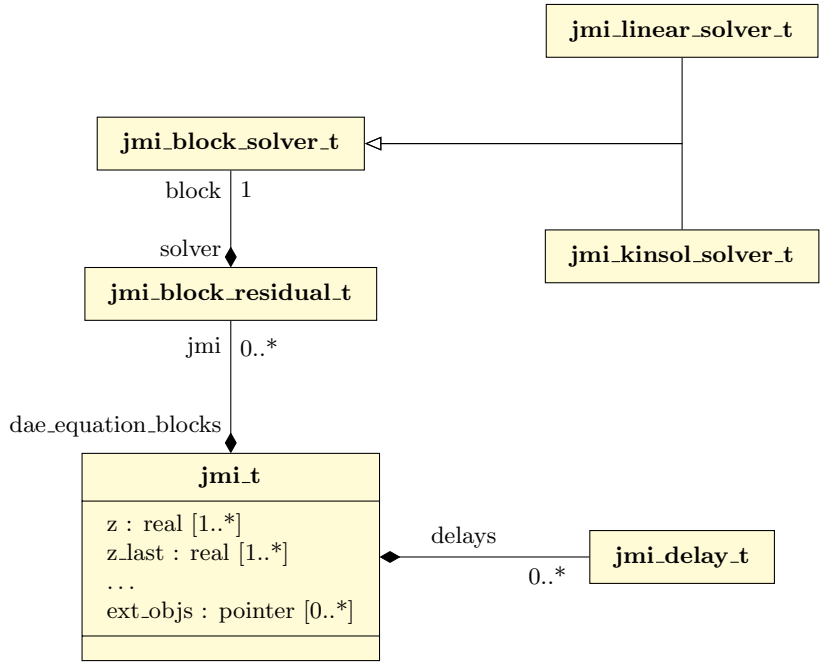

Figure 1: Conceptual Unified Modelling Language (UML) class diagram of JModelica FMU implementation.

*3.2. Enable saving and restoring of JModelica FMU internal state*

JModelica generates an FMU starting from a model written in Modelica. To do so, Modelica equations are translated into C code and compiled to produce an FMU. The output FMU consists also of a library (the same for all models) implementing the functions of the FMI standard. In JModelica, such a library is called *RuntimeLibrary*, and contains also relevant data structures responsible for the internal FMU state.

The internal FMU state is a snapshot containing all the information needed to simulate the FMU starting from the moment when the state was retrieved.

Figure 1 depicts the conceptual UML class diagram of such FMU internals focusing on the components of the internal FMU state. The main class modelling an FMU instance is `jmi_t`. The internal state of a JModelica FMU is composed of the following components.

**Values of model variables.** `jmi_t` contains several arrays of real values (*e.g.*, `z` and `z_last`) to store the information related to the actual values of model variables.

**Delay buffers.** Modelica equations can contain occurrences of `delay(expr,t,tmax)` (*delay operator*) that return `expr(time-t)`, *i.e.*, the value of the expression `expr` t time units in the past. JModelica implements such an operator by maintaining a buffer (`jmi_delay_t`) containing past values of expression `expr` from current time back to `time-t`. the optional argument `tmax` guarantees an upper bound for the buffer size when `t` is not constant during simulation. The buffer size increases until the current time is greater than `t` (`tmax` when `t` is variable). After that, the size remains constant.

**Internal state of algebraic solvers.** Algebraic equations are split into blocks (`jmi_block_residual_t`) that can be solved independently. Each such block is equipped with an instance of a linear (`jmi_linear_solver_t`) or a non-linear (`jmi_kinsol_solver_t`) solver. These solvers are stateful, so their internal state also makes part of the FMU state.

**Internal state of external objects.** Modelica allows to call external (*e.g.*, defined in C) functions (array `ext_objs` of function pointers), that can have their own state. Note that there is no way to retrieve the state of such external objects, since they are completely opaque. For example, Modelica Standard Library (MSL) has the CombiTimeTable block that computes its output signal by interpolation in a table. It is implemented using C functions that read a table from a file into an in-memory data structure and allow to query its values. From the JModelica FMU *RuntimeLibrary* perspective, such data structure is just an opaque pointer and there is no way to access it.

Note that, since the above data structures are static, FMU states have a constant size (in bytes).

We extend JModelica FMU *RuntimeLibrary* by suitably implementing the following FMI functions:

- `fmi2GetFMUstate()`. This method retrieves the current state of a given FMU as an in-memory data structure, which is opaque to the user. We refer to this method in the text as `get()`.

- `fmi2SetFMUstate()`. This method replaces the current state of an FMU with a given FMU state previously retrieved using `fmi2GetFMUstate()`. We refer to this method in the text as `set()`.

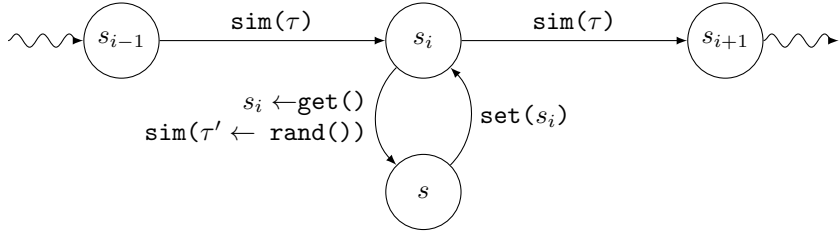

Figure 2: A glimpse of our strategy to evaluate correctness.

- `fmi2FreeFMUstate()`. This method frees up the memory occupied by a given FMU state retrieved using `fmi2GetFMUstate()`.

- `fmi2SerializeFMUstate()`. This method serializes the FMU state retrieved using `fmi2GetFMUstate()` into a byte array that can be stored in a file or sent over the network.

- `fmi2DeserializeFMUstate()`. This method deserializes a given byte array into an FMU state that can be then passed to `fmi2SetFMUstate()`.

- `fmi2SerializedFMUstateSize()`. This method returns the size of a byte array that is enough to serialize a state of the current FMU.

*3.3. Correctness evaluation strategy*

To validate our methods to get and set complete FMU states, we have to check that a call to `set()` correctly replaces the current FMU state with an FMU state obtained by a previous call to `get()`. To do so, we need to verify that further simulations are correctly computed accordingly to the actual FMU state.

In particular, we define the following two ways of performing a simulation of an input FMU.

Given an FMU and a value $\tau \in \mathbb{R}_{\geq 0}$, let $s$ be the current state of the input FMU, we denote with $s' = \text{sim}(\tau)$ a function that simulates the input FMU advancing its current state $s$ for $\tau$ time units and reaching the state $s'$.

Given an FMU and values $\tau \in \mathbb{R}_{\geq 0}$, $\tau' \in \mathbb{R}_{\geq 0}$, let $s$ be the current state of the input FMU, we denote with $s' = \text{sim}^\star(\tau, \tau')$ a function that executes the following instructions:

1. $s \leftarrow \text{get}()$;
2. $\text{sim}(\tau')$;
3. $\text{set}(s)$;
4. $s' \leftarrow \text{sim}(\tau)$.

For the sake of readability and without loss of generality, we omit the call to `fmi2FreeFMUstate()` in the above instructions as its correctness does not affect our validation process.

Intuitively, a call to $\text{sim}^\star(\tau, \tau')$, for any $\tau' \in \mathbb{R}_{\geq 0}$, is semantically equivalent to a call to $\text{sim}(\tau)$, since they both advance the current FMU state for $\tau$ units of time. This leads to the following remarks.

**Remark 1.** *Given an FMU and a value $\tau \in \mathbb{R}_{\geq 0}$ let $s$ be the current state of the input FMU. Then for all $\tau' \in \mathbb{R}_{\geq 0}$, $\text{sim}(\tau)$ and $\text{sim}^\star(\tau, \tau')$ reach exactly the same state, formally $\text{sim}(\tau) = \text{sim}^\star(\tau, \tau')$*

**Remark 2.** *Given an FMU and a value $\tau \in \mathbb{R}_{\geq 0}$, let $s$ be the initial state of the input FMU, let $\mathbf{s}_\tau = s_0, s_1, s_2, \ldots, s_n$ and $\mathbf{s}_\tau^\star = s_0^*, s_1^*, s_2^*, \ldots, s_n^*$ be the sequences of FMU states reachable from $s$ by consecutive executions of $\text{sim}(\tau)$ and $\text{sim}^\star(\tau, \tau')$, respectively, for all $\tau' \in \mathbb{R}_{\geq 0}$. If both get() and set() are implemented correctly then, for all $i \in \{0, \ldots, n\}$, $s_i = s_i^*$.*

To experimentally verify the above statement, we need to check that our implementation works for all values of $\tau'$. This is of course impossible. To overcome such an obstacle we employ a statistical approach based on hypothesis testing where $\tau'$ takes values in a bounded interval $\mathbb{B} \subset \mathbb{R}_{\geq 0}$ (see, *e.g.*, [58, 59, 60]).

Given a value for $\varepsilon \in (0, 1)$, we define a *null hypothesis* $H_0$ which states that the probability of sampling $\tau' \in \mathbb{B}$ such that the state reached by executing $\text{sim}(\tau)$ is different from the state reached by executing $\text{sim}^\star(\tau, \tau')$ is greater than $\varepsilon$. Formally $H_0$ is defined as: $H_0 : \Pr\{\tau' \in \mathbb{B} \,.\, \text{sim}(\tau) \neq \text{sim}^\star(\tau, \tau')\} \geq \varepsilon$.

Then we apply statistical hypothesis testing [61] and try to reject $H_0$ through a given number of trials $N$.

At each trial we randomly sample a value of $\tau' \in \mathbb{B}$ according to a uniform distribution and we check whether the state reached by $\text{sim}(\tau)$ is different from the state reached by $\text{sim}^\star(\tau, \tau')$ or not.

If within $N$ trials we find a value $\tau'$ such that $\text{sim}(\tau) \neq \text{sim}^\star(\tau, \tau')$, then $\tau'$ is a counterexample showing that our implementation is not correct. Formally, we prove $H_0$ when it holds.

On the other hand, rejecting $H_0$ after $N$ trials, even if it holds, introduces a *Type-I Error*.

In particular, given a value $\delta \in (0, 1)$, the probability that we make an error by rejecting $H_0$ when it holds is bounded by $\delta$.

This is shortly stated by saying that $H_0$ is rejected with statistical confidence $1 - \delta$. Finally, we conclude that the probability to sample $\tau'$ such that the state reached by executing $\text{sim}(\tau)$ is different from the state reached by executing $\text{sim}^\star(\tau, \tau')$ is less than $\varepsilon$, formally $\Pr\{\tau' \in \mathbb{B} \,.\, \text{sim}(\tau) \neq \text{sim}^\star(\tau, \tau')\} < \varepsilon$.

By exploiting results of [58], given $\delta \in (0, 1)$ and $\varepsilon \in (0, 1)$, the number of trials is computed as $N = \lceil \log(\delta) / \log(1 - \varepsilon) \rceil$.

Algorithm 1 describes our *Hypothesis Testing* approach and Figure 2 sketches our correctness evaluation strategy.

Note that at Line 6 of Algorithm 1, if, for a given $\tau'$, $\text{sim}(\tau) \neq \text{sim}^\star(\tau, \tau')$ is true, we return such a $\tau'$ as a counterexample meaning that by simulating the input FMU $i$ times we reach a state proving that get() and set() are not working correctly.

```
1 function evaluateCorrectness():
    Input: FMU
    Input: τ, simulation duration
    Input: ε ∈ (0, 1)
    Input: δ ∈ (0, 1)
    Input: 𝔹 ⊂ ℝ_{≥0}
    Output: either (True, −, −) or a counterexample (False, τ', i)
2   N ← ⌈log(δ)/log(1−ε)⌉;
    /* H_0 = Pr{τ' ∈ 𝔹 . sim(τ) ≠ sim*(τ, τ')} ≥ ε                    */
3   for i ∈ [1, N] do
4       τ' ← rand(𝔹);
5       s_i ← sim(τ); s ← sim*(τ, τ');
6       if s_i ≠ s then
7           return (False, τ', i) /* H_0 is proved                    */
8       end
9   end
10  return (True, −, −) /* H_0 is rejected                             */
```
**Algorithm 1:** Hypothesis testing approach to evaluate correctness of `set()` and `get()`.

The above considerations prove the following theorem.

**Theorem 1.** *Given an FMU having `get()` and `set()` implemented, values $\varepsilon, \delta \in (0, 1)$, a bounded interval $\mathbb{B} \subset \mathbb{R}_{\geq 0}$ and a value for $\tau \in \mathbb{R}_{\geq 0}$, Algorithm 1 is such that:*

1. *it terminates in N steps, where $N = \left\lceil \frac{\log(\delta)}{\log(1-\varepsilon)} \right\rceil$;*
2. *when it returns True, with confidence $1-\delta$: $\Pr\{\tau' \in \mathbb{B} . sim(\tau) \neq sim^{\star}(\tau, \tau')\} < \varepsilon$;*
3. *when it returns False, we have a counterexample, i.e., a value for $\tau' \in \mathbb{B}$, proving that `get()` and `set()` are not correctly implemented.*

In Section 4.3 we apply Algorithm 1 on different FMUs in order to validate our implementation of `get()` and `set()`.

*3.4. Performance evaluation strategy*

To evaluate performance of our implementation, we focus on simulation-based V&V approaches. As anticipated in Section 1, simulation-based V&V of Cyber-Physical Systems (CPSs) requires to explore all simulation scenarios. A simulation scenario is a sequence of exogenous inputs to be injected to the given CPS model, *i.e.*, FMU under verification. In this setting, the given FMU (representing the CPS under verification) is driven in the space of all possible simulation scenarios by means of a visit (see, *e.g.*, [62, 63, 64] and citations therein). The space of all simulation scenarios is defined as a tree where each edge is labelled with an exogenous input and each node represents the FMU

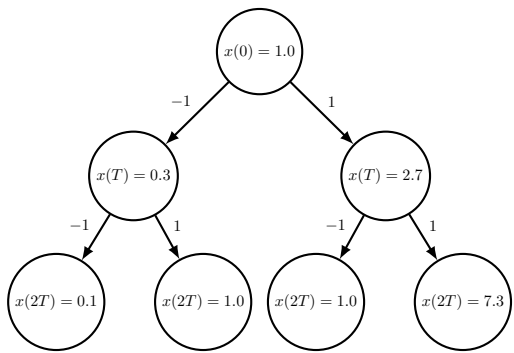

Figure 3: Example of a tree of simulation scenarios. Nodes denote FMU states whereas edges denote injected input values.

state reached by injecting the sequence of inputs associated to the path leading to that node.

As an example we can consider an FMU defining a simple hybrid system (see, *e.g.*, [10]) having only one state variable. In particular, let $x$ be a signal (i.e., a real-valued function of time) and, given a positive real number $T$ (time step), let $u$ be $T$-piecewise constant signal, i.e., a signal changing its value only at time instants of the form $kT$ ($k = 0, 1, 2, 3, \ldots$). In the following we assume that $u$ takes values in the set $\{-1, 1\}$ (i.e., for all $t$, $u(t) \in \{-1, 1\}$). When writing equations, as usual, we will write $x$ for $x(t)$ and $\dot{x}$ for $\dot{x}(t)$.

Given signals $x$ and $u$ as above, we define the behaviour of our simple example of hybrid system by means of the following differential equation:

$$\dot{x} = \begin{cases} -x & \text{if } u \text{ is } -1 \\ x & \text{if } u \text{ is } 1 \end{cases}$$

Let $x(0) = 1$ be the initial state of this FMU and $2T$ the simulation horizon. Figure 3 depicts all possible simulation scenarios starting from such an initial state and illustrates all state traversed by the FMU when the input function $u(t)$ can change value (i.e., at each $kT$ with $k = 0, 1, 2$). Note that, as described above, states reached by different sequence of inputs (tree paths) are different for us.

Typically, the goal of a simulation-based V&V activity is to search for an input sequence driving the system to an *undesirable* (error) state (e.g., to verify a safety property stating that *nothing bad* ever happens) or to a *desirable* state (e.g., to verify a liveness property stating that *something good* sooner or later will happen). Checking a liveness property typically requires looking at the future evolution of the system. As a result, in general, liveness properties cannot be casted as safety properties (see, e.g., [65]). However, in a bounded horizon setting, such as the one in our simulation-based setting, liveness properties can be casted as safety properties (see, e.g., [66, 67, 68]) stating that a given state (desirable or undesirable) is reachable, with some suitable input, within the given time horizon. Accordingly, w.l.o.g., we can cast simulation-based V&V as the problem of finding an input sequence driving the system to a given state within the given time horizon.

If such a state is not found, the time horizon is increased until an *undesirable* or *desirable* state is found, or some upper bound is reached, namely, the provided simulation horizon $h$ (see, *e.g.*, [69, 70]). This is equivalent to search for the shortest sequence of exogenous inputs that leads to a given state (see, *e.g.*, [71] and [42, 43, 72] in a finite state context) in the space of all simulation scenarios of length $h$. The simulation horizon $h$ is typically set to a value large enough to guarantee that, with high confidence, the *undesirable* or *desirable state* (if any) is reachable with an input sequence of length at most $h$. To keep $h$ small a Breadth-First Search (BFS) is used (*e.g.*, as in bounded model checking [73]). Furthermore, as anticipated in Section 1, such simulation-based V&V approaches are similar to several planning and optimisation approaches such as, *e.g.*, [74, 75, 76, 77, 31], where simulation scenarios and exogenous inputs correspond, respectively, to plans and actions to be taken in order to drive the system into a goal state.

In this setting, in order to decrease the number of simulations and, in turn, the computation time of the V&V activity (or the optimisation task), it is important to avoid simulating many times prefixes common to different simulation scenarios. To do so, it is crucial that the given FMU has the save-and-restore state functionality implemented as FMU states can be saved in memory in order to be restored later as initial states of the simulation.

Hence, such an exploration visit in the tree of simulation scenarios is a demanding application to evaluate our proposed implementation. In particular, we distinguish two kind of approaches. The former uses FMUs that are not equipped with the save-and-restore feature of FMI 2.0. We refer to this approach in the text as *without-save-restore visit*. The latter is a *save-restore visit* that uses our generated FMUs implementing those FMI features to save and restore FMU states.

It is worth noting that during a *save-restore visit*, since the state space to explore can be huge, to keep in memory the whole state space can be infeasible. To this end, several solutions have been devised in the literature (see, *e.g.*, [44, 45, 25]). The general idea is to dynamically choose the best states to store and those to forget (compatibly with memory constraints). Also, whenever no further inputs can be injected from the current node of the tree, the corresponding simulator state can be removed from memory. We refer the readers to those references for more details and algorithms that efficiently satisfy memory constraints during the visit. Accordingly, here we focus on evaluating performance of our implementation of the save-and-restore feature in terms of simulation time.

The remainder of this section is organised as follows. First, we define the space of simulation scenarios of a given CPS model (*i.e.*, FMU under verification) as a tree (Section 3.4.1). Second, we show that the time needed to explore such a tree (for both a *without-save-restore visit* and a *save-restore visit*) strictly depends on the time needed to drive the input FMU through each node of the tree (Sections 3.4.2 and 3.4.3). Last, we quantify the speed-up of a *save-restore visit* (Section 3.4.4).

*3.4.1. Tree definition*

In our setting, we perform a visit of a balanced finite tree of depth $h > 1$, where each tree node at depth $0 < i \leq h$ has a constant branching factor $b > 1$. The tree root corresponds to the initial state of the given FMU, while nodes correspond to FMU states that can be reached through simulations from the initial state. The edges of the tree correspond to possible actions that can be taken during the visit, *i.e.*, different values for the FMU exogenous inputs. For example, in the tree of Figure 3, the branching factor, $b$, and the maximum depth, $h$, are both 2.

Furthermore, to simplify calculations, during the visit we assume that the simulation of the input FMU is advanced by a fixed quantity of $\tau \in \mathbb{R}_{>0}$ time units each time a new node is being visited (in our example $\tau$ corresponds to $T$, *i.e.*, 1 second). Also, to simplify our analysis, we assume that the execution time of a simulation depends on the current FMU state and on the simulation duration, *i.e.*, $\tau$, and not on values of exogenous inputs. This might not be strictly true for each simulation, but it is a very reasonable assumption on average for typical CPS models.

*3.4.2. Cost of a without-save-restore visit*

A *without-save-restore visit* aims at driving FMUs without get-and-set state capabilities. Hence, when we have to explore a node in the tree, we need to simulate the given FMU from its initial state up to the state denoted by such a node. This means that a node at depth $i > 0$ can be reached after a simulation of $i\tau$ time units from the initial state of the given FMU. We denote such operation with $\mathtt{sim}(i\tau)$. For example, in Figure 3, the state $x(2T) = 0.1$ can be reached by means of a $\mathtt{sim}(2\tau)$ operation, where $\tau = T$, and by injecting $u(0) = -1$ and $u(T) = -1$. As already stated in Section 3.3, $\mathtt{sim}()$ can be defined according to FMI 2.0 specifications.

We define our cost function as $\mathcal{C}(i) = \mathcal{T}(\mathtt{sim}(i\tau))$, which measures the execution time, $\mathcal{T}$, to reach a node at depth $i > 0$. Finally, we define the total execution time of a *without-save-restore visit* as the sum of the costs of each single node in the tree:

$$\sum_{i=1}^{h} \mathcal{C}(i) \, b^i \tag{1}$$

*3.4.3. Cost of a save-restore visit*

A *save-restore visit* aims at driving FMUs by exploiting their get and set capabilities in order to simulate only once each prefix common to different paths of the given tree. To do so, each time a node at depth $i > 0$ has to be visited, we perform the following steps:

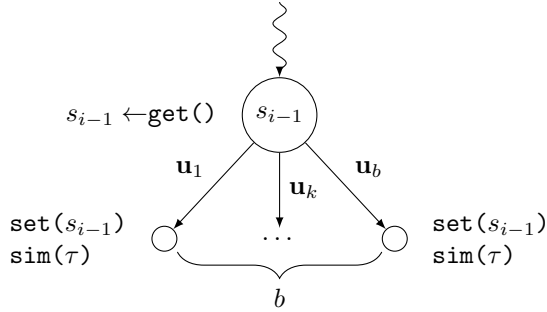

Figure 4: Usage of set() and get() to explore nodes during a *save-restore visit*

1. set(), to restore (load) the state corresponding to the node at depth $i-1$ as a start state of the given FMU;
2. sim($\tau$), to advance the current FMU state to the state reached after a simulation of $\tau$ time units also by injecting the value for exogenous inputs corresponding to the traversed edge.
3. get(), to store (save), *e.g.*, on disk, the reached FMU state for further steps.

For the sake of readability, we omit the time needed to perform a fmi2FreeFMUstate() because it is negligible. Figure 4 outlines the above steps.

As in the previous section, we define a cost function, *i.e.*, $\mathcal{C}^\star(i)$, which denotes the execution time spent to visit a node at depth $i > 0$ during a *save-restore visit*, as follows.

$$\mathcal{C}^\star(i) = \mathcal{T}(\frac{1}{b}\,\texttt{get()} + \texttt{set()} + \texttt{sim}(\tau)).$$

Note that the get() is performed at depth $i-1$ (see Figure 4) as the corresponding retrieved FMU state will be then used as a start state for all the children of that node. However, to simplify the formulation we count such get() time in the child nodes. Hence, each child node (at depth $i$) contributes $\frac{1}{b}$get() to the total get() time, *i.e.*, the total get() time is amortised by the branching factor.

Accordingly, along the lines of Eq. (1), the total execution time needed by a *save-restore visit* to explore the whole tree space is equal to the sum of the costs of each node:

$$\sum_{i=1}^{h} \mathcal{C}^\star(i)\, b^i \qquad (2)$$

*3.4.4. Speed-up of a* save-restore visit

We define the speed-up, $S(h,b)$, as the ratio between the cost of a *without-save-restore visit*, Eq. (1), and the cost of a *save-restore visit*, Eq. (2), on a perfectly balanced tree having branching factor $b$ and depth $h$:

$$S(h,b) = \frac{\sum_{i=1}^{h} \mathcal{C}(i)\, b^i}{\sum_{i=1}^{h} \mathcal{C}^\star(i)\, b^i} \qquad (3)$$

When the above ratio is greater than 1 it means that a *save-restore visit* is faster than a *without-save-restore visit*.

Furthermore, we note that the above formula, *i.e.*, Eq. (3), can be simplified under the following assumptions.

First, the time spent for simulating an FMU for $i\tau$ time units, *i.e.*, $\mathcal{T}(\texttt{sim}(i\tau))$, is approximately equal to $i \times \overline{\mathcal{T}}(\texttt{sim}(\tau))$, that is $i$ times the average execution time of a simulation of length $\tau$.

Second, we can use average execution times of $\texttt{get()}$ and $\texttt{set()}$, *i.e.*, $\overline{\mathcal{T}}(\texttt{get()})$ and $\overline{\mathcal{T}}(\texttt{set()})$, respectively. As Section 3.2 describes, this is motivated by the fact that states of a given FMU have a constant size (in bytes).

In Section 4.4, we experimentally show that such assumptions are reasonable.

We rewrite $\mathcal{C}(i)$ as $\mathcal{C}(i) = i \times \overline{\mathcal{T}}(\texttt{sim}(\tau))$ and $\mathcal{C}^\star(i)$ as $\mathcal{C}^\star(i) = \frac{1}{b}\,\overline{\mathcal{T}}(\texttt{get()}) + \overline{\mathcal{T}}(\texttt{set()}) + \overline{\mathcal{T}}(\texttt{sim}(\tau))$.

Hence, the speed-up of Eq. (3), $S(h,b)$, can be written as follows:

$$S(h,b) = \frac{\overline{\mathcal{T}}(\texttt{sim}(\tau)) \sum_{i=1}^{h} i b^i}{\left(\frac{1}{b}\,\overline{\mathcal{T}}(\texttt{get()}) + \overline{\mathcal{T}}(\texttt{set()}) + \overline{\mathcal{T}}(\texttt{sim}(\tau))\right) \sum_{i=1}^{h} b^i}.$$

This leads to the following proposition.

**Proposition 1.** *Given* $\mathcal{C}(i) = i \times \overline{\mathcal{T}}(\textit{sim}(\tau))$ *and* $\mathcal{C}^\star(i) = \frac{1}{b}\,\overline{\mathcal{T}}(\textit{get()}) + \overline{\mathcal{T}}(\textit{set()}) + \overline{\mathcal{T}}(\textit{sim}(\tau))$, *the speed-up of a* save-restore visit *is greater than 1, i.e., $S(h,b) > 1$, when the following equation is satisfied:*

$$\frac{\frac{1}{b}\,\overline{\mathcal{T}}(\textit{get()}) + \overline{\mathcal{T}}(\textit{set()})}{\overline{\mathcal{T}}(\textit{sim}(\tau))} < \psi(h,b) \qquad (4)$$

*where $\psi(h,b)$ is a threshold defined as:*

$$\psi(h,b) = \frac{\sum_{i=1}^{h} i b^i}{\sum_{i=1}^{h} b^i} - 1 = \frac{h b^{h+1} - (h+1) b^h + 1}{(b-1)(b^h - 1)} - 1. \qquad (5)$$

**Remark 3.** *When $b$ is very large, formally $b \to \infty$, and $h > 1$, the speed-up threshold $\psi(h,b)$ approaches to $h-1$, formally $\psi(h,b) \to h-1$. Thus, Proposition 1 can be rephrased saying that $S(h,b) > 1$ when the following condition is satisfied.*

$$R(h,\tau) = \frac{\overline{\mathcal{T}}(\textit{set()})}{(h-1)\overline{\mathcal{T}}(\textit{sim}(\tau))} < 1. \qquad (6)$$

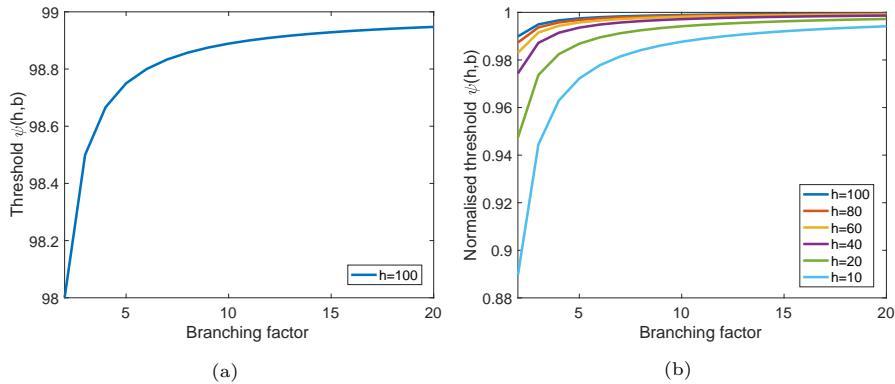

Figure 5: Threshold $\psi(h, b)$ computed for $h = 100$ (a) and for different values of $h$ (b). In the right figure (b), our threshold is also normalised by $h - 1$ for presentation purposes.

Intuitively, Remark 3 says that when the branching factor of a given tree is sufficiently large and the ratio $R(h, \tau)$ is less than 1, we expect that the use of get/set state functionality speeds up a *without-save-restore visit*. Note that, since the get() is executed once for each node of the tree and the retrieved state is then used as many times as the value of the branching factor $b$ of that node (*i.e.*, for each child node), the inefficiency of a get() is amortised among those child nodes. Hence, even if for some model the average time of a get() is high because the operation is inefficient, the term $\frac{1}{b}$get() of Eq. (4) becomes negligible. In Section 4.4.1 we show that our assumptions that motivate this remark are reasonable for a large set of benchmark models.

In Figure 5a we show that, starting from small values of $b$, the threshold $\psi(h, b)$ rapidly reaches values towards the limit $h - 1$. This is true also for different values of $h$ as Figure 5b shows.

Hence, under our assumptions, given an FMU and a tree having a low branching factor, namely 4-5, if the average execution time needed to perform a set() operation is lower than the time to perform a simulation of a scenario of length $h - 1$, then the speed-up of a *save-restore visit* with respect to a *without-save-restore visit* is greater than 1.

In Section 4.4, we present experimental results regarding the speed up computation on different FMUs by taking into account trees of different depths and branching factors.

## 4. Experimental Results

In this section, we present experimental results to evaluate the correctness and the effectiveness of our implementation against state-of-the-art case studies. After briefly describing such case studies, we evaluate correctness and we finally show how our implementation enhances performance of simulation-based Verification and Validation (V&V) approaches.

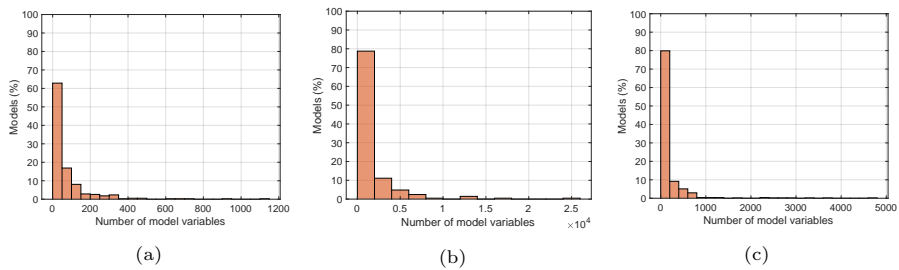

Figure 6: Distribution of sizes of models within MSL (a), STS (b) and BMD (c).

## 4.1. Case studies

In Section 4.4 we see that, depending on the Functional Mock-up Unit (FMU) at hand, different components of the FMU architecture are involved during saving and restoring an FMU state.

To assess correctness and to evaluate performance on all such components we evaluate our implementation by using FMUs generated from the following widely established model libraries.

**Modelica Standard Library (MSL):** the official Modelica library collecting models that are developed and reviewed by the Modelica Association (version 3.2.2, https://modelica.org). It provides models and components from different engineering domains such as mechanical, electrical, magnetic, fluid, thermal and control systems. The library contains both models defining standardised interfaces or building blocks and models that are directly usable. For our purpose, we focus on the latter class of models, which comprises 385 models having 150 model variables on average.

**Scalable Test Suite (STS):** a Modelica library of benchmark models useful for assessing performance of large scale systems (version 1.11.4, see [78]). The library contains 16 models covering electrical, mechanical, power and thermal domains. Such models are scalable in terms of their size (*i.e.*, number of variables). The STS provides also 207 ready-to-run models validated by the authors and having an average of 2913 model variables.

**BioModels Database (BMD):** a well-known repository of mathematical models of biological systems taken from the scientific literature [79]. A subset of these models, consisting of manually curated models, is widely-used as a benchmark for Systems Biology Markup Language (SBML) simulators. From such a subset, we used models already translated from SBML to Modelica and validated in [80]. In total we consider 411 models having 370 model variables on average.

Figure 6 shows the distribution of model sizes in terms of the number of model variables.

*4.2. Experimental setting*

All our experiments have been carried out on a High Performance Computing (HPC) infrastructure (*i.e.*, Marconi cluster at CINECA, Italy).

For each Modelica model within our datasets, *i.e.*, MSL, STS and BMD, we generated its FMU using our extended JModelica.

We manually excluded from our datasets those FMUs that are not compatible with get/set state functionality. These are FMUs that use external objects (as described in Section 3.2). This led us to exclude 49 (*i.e.*, 12.73%) models from MSL and 20 (*i.e.*, 9.66%) models from STS. All models within BMD have been included. Hence, in total, we excluded the 6.88% of all FMUs within our datasets.

In order to assess correctness and evaluate performance of our implementation of get/set functionality, FMUs have been simulated using `SUNDIALS CVODE` [81] solver by means of the `PyFMI` library [82]. For each FMU, we set the value of $\tau$ to 1% of the FMU default simulation horizon, *i.e.*, the value of FMU `default experiment stop time`.

*4.3. Correctness evaluation results*

Correctness has been evaluated by means of the approach presented in Section 3.3. In particular, for each of the 934 FMUs, we ran our Algorithm 1 to assess that our implementation correctly restores a previously-saved FMU state.

To do so, we consider as $\mathbb{B}$ values from 0 to the default simulation horizon of the given FMU and we set $\delta = 8\%$ and $\varepsilon = 2.5\%$. Thus the number of trials has been set to $N = 100$.

During the sampling process, the inequality of FMU states (Line 6 of Algorithm 1) has been checked through a bit-wise comparison. For all our 934 FMUs our Algorithm 1 returned true, proving that our implementation is correct with a degree of statistical confidence of 92%.

*4.4. Performance evaluation results*

In this section, we describe experimental results of our performance evaluation according to our strategy described in Section 3.4. In particular, in Section 4.4.1 we evaluate performance of our implementation by a preliminary analysis using results of Remark 3. Then, in Section 4.4.2 we compute the speed-up of a *save-restore visit* with respect to a *without-save-restore visit* on trees with different values of depth and branching factor.

*4.4.1. Preliminary analysis*

As a preliminary analysis, for each FMU, we computed the ratio $R(h, \tau)$ (see Remark 3) to compare the execution time of a `set()` with the execution time needed for simulating the given FMU.

Such an analysis has been conducted by taking into account different values for the depth of the tree, *i.e.*, $h \in [1, 100]$.

To perform such an analysis we show that our assumptions (*i.e.*, those described in Section 3.4.3) are reasonable for our case-study FMUs. To do so, in

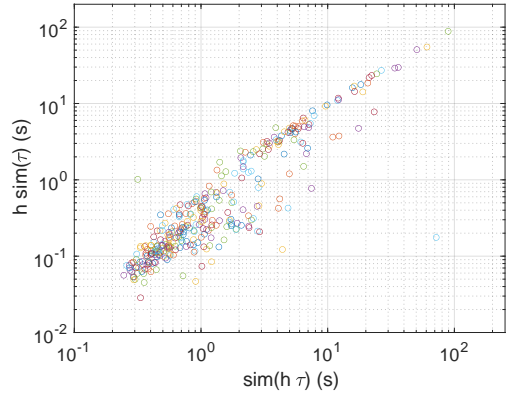

(a)

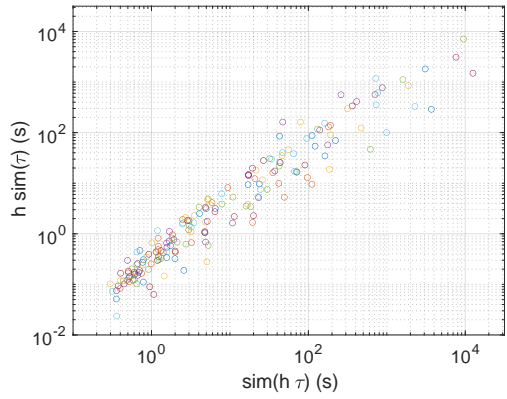

(b)

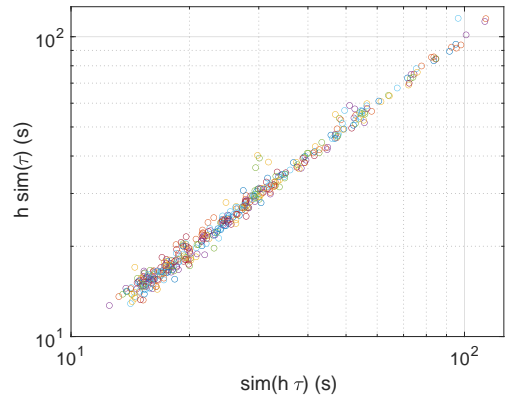

(c)

Figure 7: Correlation among the execution times (in seconds) of a simulation of duration $h \times \tau$ (on x axis) and $h$ simulations of duration $\tau$ (on y axis), where $h = 100$. Each marker represents an FMU within MSL (a), STS (b) and BMD (c).

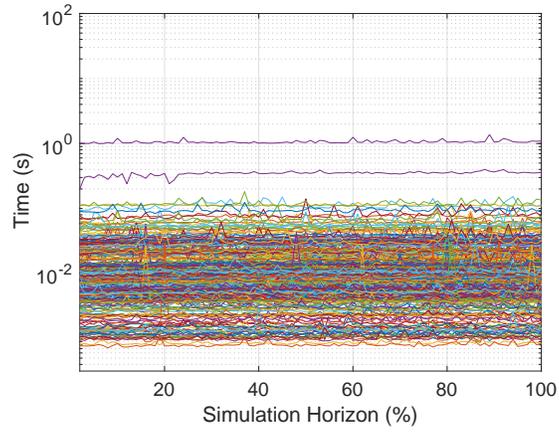

(a)

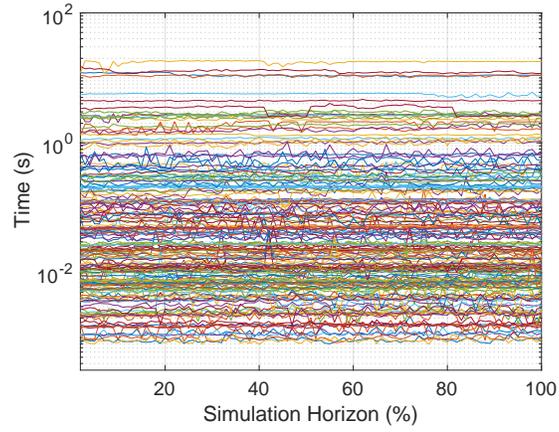

(b)

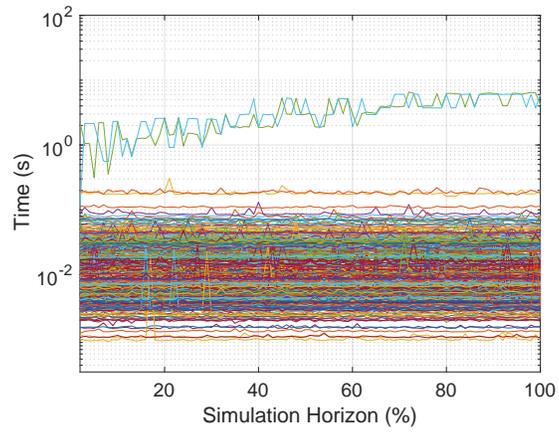

(c)

Figure 8: Execution times (in seconds) spent for performing `set()` during FMU simulations. Each line represents an FMU within MSL (a), STS (b) and BMD (c).

Figure 7, we compared the execution time of a simulation of length $i \times \tau$ with the execution time of $i$ simulations of length $\tau$. As we see, these two quantities are highly correlated. Indeed, the Pearson correlation coefficient, $\rho$, for MSL, STS and BMD is 0.88, 0.79 and 0.99, respectively.

Moreover, in Figure 8, for each FMU, we show that the execution time (in seconds) spent to perform a `set()` operation is almost constant during the FMU simulation (from its start time to its simulation horizon).

There is also an atypical behaviour for two FMUs within BMD. In such models, the `delay` operator is used and parameter `tmax` is always set to be as large as possible (much larger than the given FMU simulation horizon). This causes the size of the associated delay buffers to increase after each simulation in order to store all needed delay variable values. For this reason, the size of the internal FMU state also increases together with the execution time needed to perform `set()` (and also `get()` as well).

Having clarified that our assumptions are reasonable, in Figure 9 we show how $R(h, \tau)$ changes by varying the depth of the tree, for each FMU in our datasets.

As we expected, for the majority of the analysed FMUs, the values of $R(h, \tau)$ are either below 1 or they go below 1 for very small values of depth (namely, $< 20$). This means that the cost of `set()` is very low with respect to simulations in terms of computation time.

Also, there is one single FMU having $R(h, \tau)$ greater than 1 for all values of depth. Such behaviour is explained by the fact that the larger the state of a given FMU the slower the execution of `set()` in terms of computation time. Hence, a *without-save-restore visit* performs better for those FMUs having a very large state (thousands of variables, as for FMUs within STS) but very fast in simulating. Of course, as we described in Section 3.4.3, this is strictly linked also to the depth of the given tree which affects the length of simulations to be performed in a *without-save-restore visit*.

Note that, $R(h, \tau)$ is less than 1 also for those two FMUs having a non-constant execution time for `set()` (see Figure 8c).

*4.4.2. Speed-up analysis*

In order to perform a more in-depth analysis, we also computed our cost functions for the *without-save-restore visit*, *i.e.*, Eq. (1), and the *save-restore visit*, *i.e.*, Eq. (2), and finally our speed-up formula, *i.e.*, $S(h, b)$ (see Eq. (3)).

To do so, different values for the branching factor, *i.e.*, $b \in [2, 10]$, and for the depth of the tree, *i.e.*, $h \in [1, 100]$, have been taken into account. Note that, such chosen values are perfectly reasonable for real case studies. For example, in [83, 31] a simulation-based approach consisting in a backtracking-based search has been employed in a search space defined as a tree of depth equal to 55 and constant branching factor equal to 3.

Figure 10 shows the average speed-up computed among FMUs within MSL, STS and BMD when varying both the branching factor and the depth of the tree.

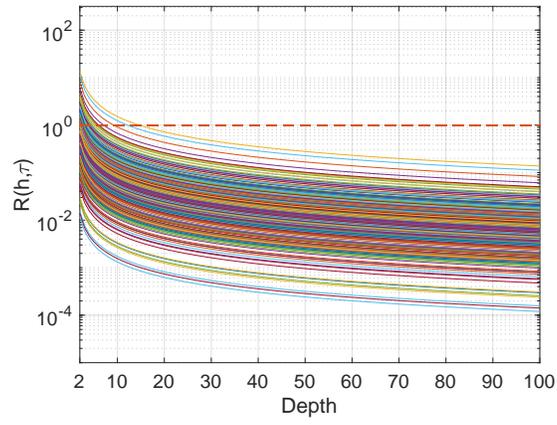

(a)

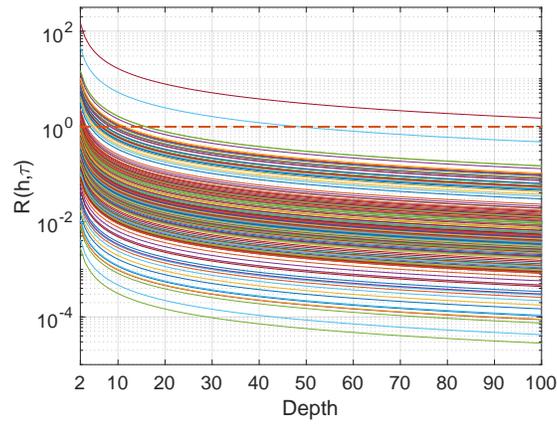

(b)

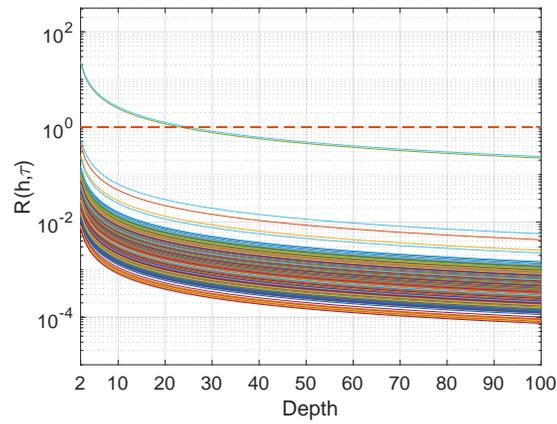

(c)

Figure 9: Computation of $R(h,\tau)$, where $h \in [1, 100]$. Dashed lines represent thresholds of $R(h,\tau)$. Each curve represents an FMU within MSL (a), STS (b) and BMD (c).

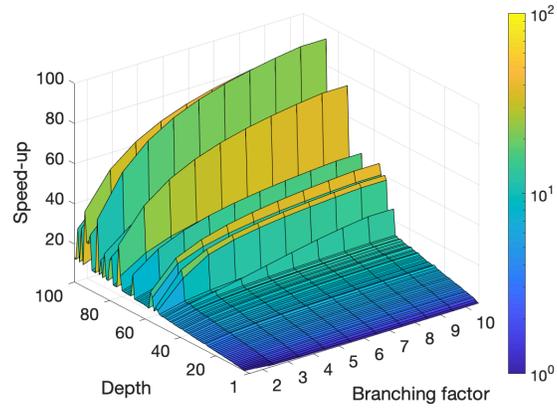

(a)

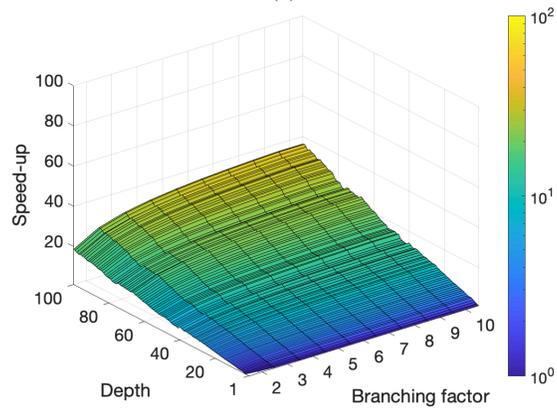

(b)

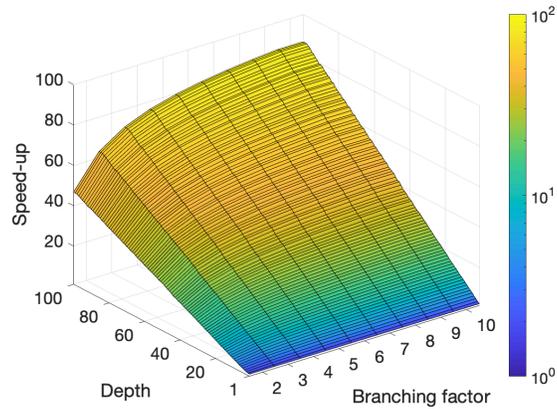

(c)

Figure 10: Average speed-up (heat map) computed among FMUs within MSL (a), STS (b) and BMD (c) by varying the branching factor and the depth of the tree.

As we expected from our preliminary analysis (Section 4.4.1), even with a low value for the branching factor, *i.e.*, 5, and a low depth, *i.e.*, 50, a *save-restore visit* is, on average, 22 times faster than a *without-save-restore visit*, among all our case studies. In particular, for FMUs within MSL, STS and BMD, we have on average a speed-up of 11.75, 14.06 and 39.84, respectively.

Furthermore, by keeping constant the given tree depth and branching factor, $h = 100$ and $b = 10$, we reach, on average, a speed-up value of 26.62, 36.11 and 86.37, among all FMUs within MSL, STS and BMD, respectively.

As we can note, FMUs within BMD (Figure 10c) achieve higher values of speed-up than FMUs within the other datasets. It is clear (starting from depth equal to 2) that the execution time of `set()` is very low with respect to simulation execution time. Hence, the deeper the tree (*i.e.*, the number of simulations) higher the speed-up achieved.

Surprisingly, as Figure 10a shows, for some values of $h$ the computed speed-up is around 70. This behaviour is due to the fact that, for some FMUs within MSL the numerical integrator reaches a computationally expensive integration step which requires more time to be solved, *e.g.*, a chattering effect.

Hence, having the FMU at hand equipped with `get()` and `set()` functionalities brings a huge benefit. Indeed, once this complex integration step is solved, the reached FMU state can be saved and restored for further simulations.

Figure 11 clearly shows this by presenting, for each evaluated FMU, the speed-up achieved when varying the depth of the tree, while keeping constant the value of the branching factor ($b = 10$). In particular, in Figure 11a we see that few FMUs reach peaks of speed-up above 10 000.

Furthermore, in both Figures 10 and 11, we note that the speed-up increases almost linearly with respect to values of the depth of the tree.

## 5. Conclusions

We have presented an extended version of JModelica that implements Functional Mock-up Interface (FMI) 2.0 methods to save and restore complete states of Functional Mock-up Units (FMUs) for Model Exchange (ME). We have conducted an in-depth evaluation of our implementation on FMUs generated from widely established model libraries, namely Modelica Standard Library (MSL), Scalable Test Suite (STS) and BioModels Database (BMD). We have shown the correctness of our proposed implementation and assessed its performance on a demanding application such as simulation-based Verification and Validation (V&V) approaches that drive the input FMU in the space of all simulation scenarios. In doing so, we computed the speed-up of a visit which drives FMUs generated by our extended JModelica implementation with respect to a visit which drives FMUs generated by stand-alone JModelica. We have shown even for a tree with a low branching factor, namely 5, and a low depth, namely 50, the speed-up achieved is, on average, 11.75, 14.06 and 39.84, among all FMUs within MSL, STS and BMD, respectively. Also, the achieved speed-up increases much more for deeper trees. Future directions can be to apply our save-and-restore implementation to FMUs for Co-Simulation (CS) as well as to port our implementation to the OpenModelica simulation environment. Moreover, as the 3.0 version of FMI standard is planned to be released, it could be worth investigating how to adapt and extend our implementation to the new set of envisioned features (*e.g.*, clocks and hybrid co-simulation to handle event-driven dynamics and a new type of model exchange format, *i.e.*, Scheduled Execution, SE).

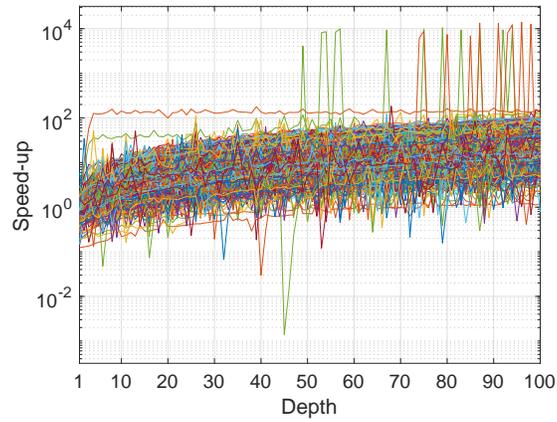

(a)

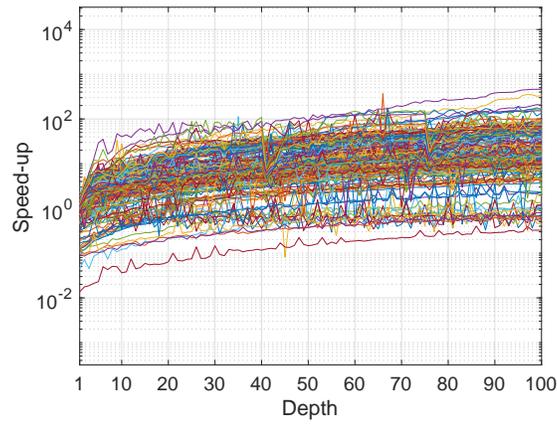

(b)

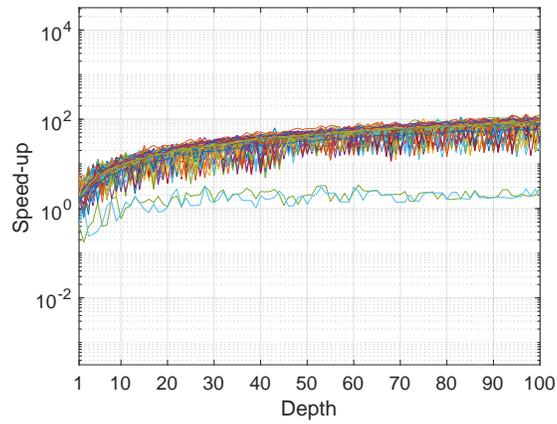

(c)

Figure 11: Speed-up achieved for each FMU within MSL (a), STS (b) and BMD (c) when the branching factor is equal to 10 and the depth takes values within $[1, 100]$.

**Acknowledgements**

The authors are grateful to their alumni, Agostina Calabrese, Michele Laurenti, and Alessandro Steri, for their contribution to develop a first prototype during their bachelor thesis work.

This work was partially supported by the following research projects and grants: Italian Ministry of University & Research (MIUR) grant "Dipartimenti di Eccellenza 2018–2022" (Dept. Computer Science, Sapienza Univ. of Rome); EC FP7 project PAEON (Model Driven Computation of Treatments for Infertility Related Endocrinological Diseases, 600773); EC FP7 project SmartHG (Energy Demand Aware Open Services for Smart Grid Intelligent Automation, 317761); Sapienza University 2018 project RG11816436BD4F21 "Computing Complete Cohorts of Virtual Phenotypes for In Silico CLinical Trials and Model-Based Precision Medicine"; A system for UAV detection (Aerospace and security,POR FESR 2014-2020); INdAM "GNCS Project 2020"; CINECA Class C ISCRA Project (HP10CPJ9LW).

**Appendix A. Proof of Results**

**Proposition 1.** *Given* $\mathcal{C}(i) = i \times \overline{\mathcal{T}}(\mathit{sim}(\tau))$ *and* $\mathcal{C}^{\star}(i) = \frac{1}{b}\overline{\mathcal{T}}(\mathit{get}()) + \overline{\mathcal{T}}(\mathit{set}()) + \overline{\mathcal{T}}(\mathit{sim}(\tau))$, *the speed-up of a* save-restore *visit is greater than 1, i.e.,* $S(h,b) > 1$, *when the following equation is satisfied:*

$$\frac{\frac{1}{b}\overline{\mathcal{T}}(\mathit{get}()) + \overline{\mathcal{T}}(\mathit{set}())}{\overline{\mathcal{T}}(\mathit{sim}(\tau))} < \psi(h,b) \qquad (4)$$

*where* $\psi(h,b)$ *is a threshold defined as:*

$$\psi(h,b) = \frac{\sum_{i=1}^{h} i b^i}{\sum_{i=1}^{h} b^i} - 1 = \frac{hb^{h+1} - (h+1)b^h + 1}{(b-1)(b^h - 1)} - 1. \qquad (5)$$

*Proof.*

$$S(h,b) = \frac{\overline{\mathcal{T}}(\mathtt{sim}(\tau)) \sum_{i=1}^{h} i b^i}{\left(\frac{1}{b}\overline{\mathcal{T}}(\mathtt{get}()) + \overline{\mathcal{T}}(\mathtt{set}()) + \overline{\mathcal{T}}(\mathtt{sim}(\tau))\right) \sum_{i=1}^{h} b^i}.$$

Requiring that $S(h,b) > 1$, we have the following threshold:

$$\frac{\frac{1}{b}\overline{\mathcal{T}}(\mathtt{get}()) + \overline{\mathcal{T}}(\mathtt{set}())}{\overline{\mathcal{T}}(\mathtt{sim}(\tau))} < \frac{\sum_{i=1}^{h} i b^i}{\sum_{i=1}^{h} b^i} - 1$$

Given the fact that: $\sum_{i=1}^{h} i b^i = b \frac{\mathrm{d}}{\mathrm{d}b}\left(\sum_{i=1}^{h} b^i\right)$ we solve summations in the following way:

$$\sum_{i=1}^{h} i b^i = b \frac{\mathrm{d}}{\mathrm{d}b} \sum_{i=1}^{h} b^i = b \frac{\mathrm{d}}{\mathrm{d}b}\left[\frac{b(b^h - 1)}{b - 1}\right] = \frac{b(hb^{h+1} - (h+1)b^h + 1)}{(b-1)^2}$$

This leads to Eq. (4):

$$\frac{\frac{1}{b}\overline{\mathcal{T}}(\mathtt{get}()) + \overline{\mathcal{T}}(\mathtt{set}())}{\overline{\mathcal{T}}(\mathtt{sim}(\tau))} < \frac{hb^{h+1} - (h+1)b^h + 1}{(b-1)(b^h - 1)} - 1$$

$\square$

**Remark 3.** *When* $b$ *is very large, formally* $b \to \infty$, *and* $h > 1$, *the speed-up threshold* $\psi(h,b)$ *approaches to* $h - 1$, *formally* $\psi(h,b) \to h - 1$. *Thus, Proposition 1 can be rephrased saying that* $S(h,b) > 1$ *when the following condition is satisfied.*

$$R(h,\tau) = \frac{\overline{\mathcal{T}}(\mathit{set}())}{(h-1)\overline{\mathcal{T}}(\mathit{sim}(\tau))} < 1. \qquad (6)$$

*Proof.* We note that, when $b \to \infty$ the left hand side of Eq. (4) converges to:

$$lim_{b\to\infty} \frac{\frac{1}{b}\overline{\mathcal{T}}(\texttt{get()}) + \overline{\mathcal{T}}(\texttt{set()})}{\overline{\mathcal{T}}(\texttt{sim}(\tau))} = \frac{\overline{\mathcal{T}}(\texttt{set()})}{\overline{\mathcal{T}}(\texttt{sim}(\tau))}$$

while, the right hand side of Eq. (4) converges to:

$$lim_{b\to\infty} \frac{hb^{h+1} - (h+1)b^h + 1}{(b-1)(b^h - 1)} - 1 = h - 1$$

Thus, when $h > 1$, our threshold is simplified in the following way:

$$\frac{\overline{\mathcal{T}}(\texttt{set()})}{(h-1)\overline{\mathcal{T}}(\texttt{sim}(\tau))} < 1$$

□

**Appendix B. Software**

Our extended JModelica 2.1 is publicly available at https://bitbucket.org/mclab/jmodelica.org

Software tools used for evaluating correctness and performance of our implementation are available at https://bitbucket.org/mclab/getsetfmustateeval